\documentclass[10pt]{iopart}

\usepackage{graphicx}

\usepackage{hyperref} 

\newcommand{\beq}{\begin{equation}}
\newcommand{\eeq}{\end{equation}}

\newcommand{\ben}{\begin{enumerate}}
\newcommand{\een}{\end{enumerate}}

\begin{document}

\title[Cuprates]{Review cuprates -- Anderson's unhappy electrons and their fate}

\author{Navinder Singh}

\address{Theoretical Physics Division, Physical Research Laboratory, Ahmedabad. India. PIN: 380009.}
\ead{navinder.phy@gmail.com; navinder@prl.res.in}
\vspace{10pt}
\begin{indented}
\item[]7/7/2023
\end{indented}

\begin{abstract}
In cuprates, as doping $p$ is reduced from the overdoped side through the quantum critical point $p^*$, a transition in Hall number density of carriers is observed, in which this number of carriers reduces from $1+p$ holes per copper site to $p$ holes per copper site. The connection of this $1+p$ to $p$ transition with pseudogap and with superconductivity is discussed. A "panoramic" view or a "broad-brush" discussion is presented in which Anderson's "unhappy" electrons take on a central stage.
\end{abstract}

\vspace{2pc}


\ioptwocol

\noindent
{"He [Bardeen] felt that formalism could lead one astray, unless it was closely tied to experimental and physical intuition" --Schrieffer on Bardeen.
\vspace{5pc}

P. W. Anderson (Figure 1) while discussing the crystal and electronic structure of cuprates argued that both localized (magnetic) and mobile (itinerant) degrees-of-freedom reside in the $CuO_2$ planes of lamellar cuprate high-$T_c$ superconductors\cite{ander1}. Spacer layers separating the $CuO_2$ planes act like charge reservoirs. By changing the chemical composition of these spacer layers hole doping or electron doping in $CuO_2$ planes can be achieved. 

\begin{figure}[!h]
\begin{center}
\includegraphics[height=10cm]{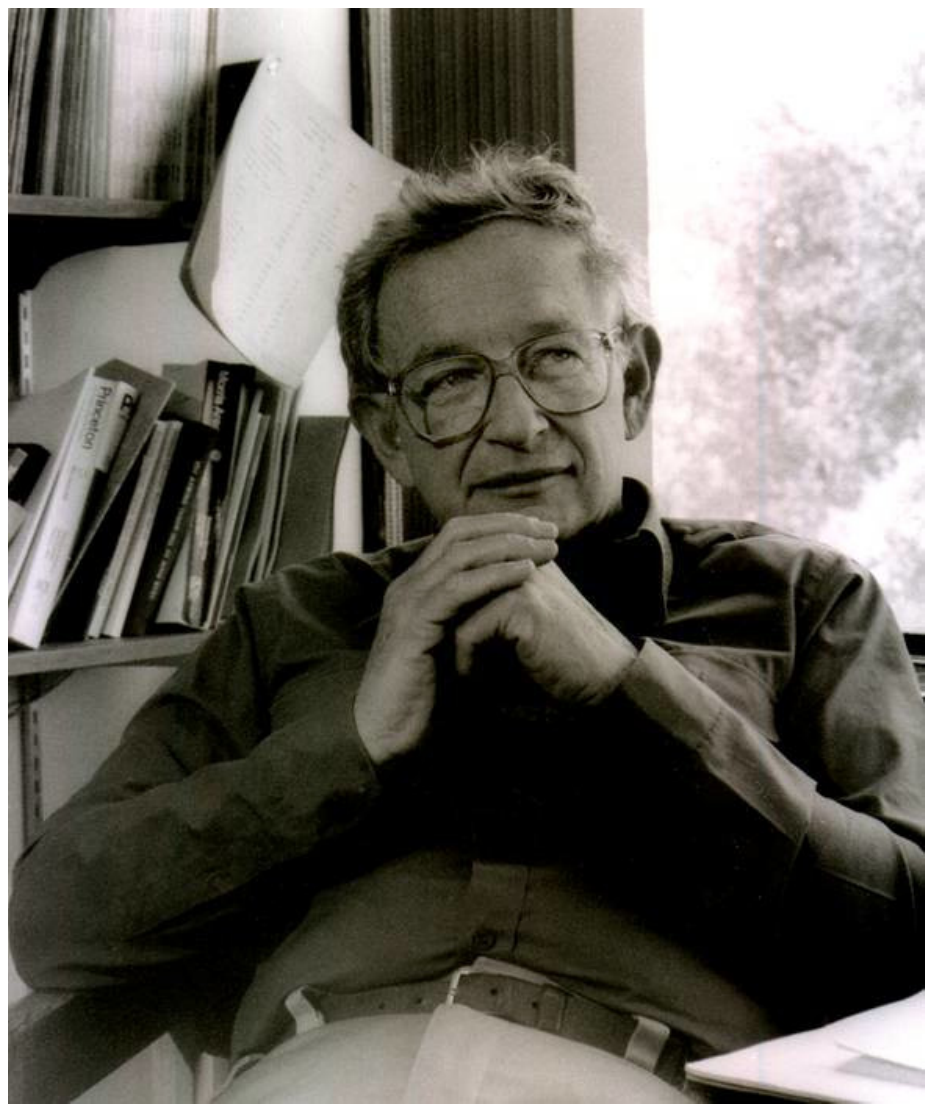}
\caption{Author pays his tribute to P. W. Anderson (December 13, 1923 -- March 29, 2020) for contributing some of the pioneering ideas, right way to approach the problem, and for his very insightful articles on the subject, especially\cite{ander1}.}
\end{center}
\end{figure}

Simple valence counting shows that in the undoped compounds (for example in $La_2CuO_4$) copper atoms should have the nominal valence state of nine electrons in the valence d-orbitals ($Cu^{++}: Ar[3d^9]$). He further argued that due to crystal field effects (created by an octahedron geometry of the cages of oxygen atoms around copper atoms) copper d orbitals of the planner copper atoms in $CuO_2$ planes of cuprates split into two sub-groups $e_g$ and $t_{2g}$,  whereas $t_{2g}$ forms a lower energy set that is filled first and $e_g$ form  a higher energy set that is filled last. Out of nine electrons in the relevant copper $3d^9$ valence state, six electrons are filled first in $t_{2g}$ orbitals i.e., in $d_{xy},~d_{yz},~d_{zx}$ orbitals.  The remaining three electrons go into the $e_g$ set which contains the $d_{x^2-y^2}$ and $d_{3z^2-r^2}$ orbitals. It turns out that the orbitals $d_{3z^2-r^2}$ which have lobs perpendicular to the $CuO_2$ planes and pointing towards the apical oxygen atoms in the octahedron geometry has lower energy as compared to that of  $d_{x^2-y^2}$ and two electrons are filled in it.\footnote{This splitting in energy between $d_{x^2-y^2}$ and $d_{3z^2-r^2}$ orbitals is due to Jahn-Teller distortions of the octahedron cage. Apical oxygen atoms are pushed out whereas planner ones are pulled in making the octahedron pointy, especially in LSCO\cite{ander1}.}  The remaining electron goes into the highest energy $d_{x^2-y^2}$ orbital which has four lobs pointing towards oxygen atoms in the $CuO_2$ plane. Anderson called the lobs pointing towards planner oxygen atoms as unhappy lobs, and this lone (un-paired) electron is an "unhappy electron".  "Unhappy", because of its highest energy state. Thus there is a hole and an electron in the highest energy $d_{x^2-y^2}$ orbital. {\it In these "unhappy" electrons there lies the secret of cuprates}. It is the seat of magnetic degrees of freedom at zero doping and at finite doping too (although long range magnetic order is "melted" at a finite doping of about 0.05). It is also the seat of mobile degrees of freedom at finite doping as the system becomes conducting. It exhibits dual character (completely localized at zero doping and quasi-localized at any finite doping but below overdoping, more precisely below $p^*=0.19$ as discussed in detail below).  In a more accurate description d-p hybridization between copper d orbitals and oxygen p orbitals leads to bonding and anti-bonding sets of orbitals, and it turns out that the hybrid $Cu3d_{x^2-y^2}--O_{2p\sigma}$ orbitals constitute higher energy states and form a narrow-hybrid-d-p band in which "unhappy" electrons reside and where electronic correlations play a dominant role.

In the following paragraphs, we sketch a physical picture in which the behaviour of these "unhappy" electrons as a function of doping and temperature takes a central stage.

These "unhappy" electrons are localized at zero doping (due to strong on-site repulsion $U$) and exhibit AFM order (local moments form at $T\sim J/k_B \sim 1200~K$, but, full 3-D AFM order (via inter-layer interactions) sets at a much lower temperature, at around $300~K$).  When the system is doped with holes (or when the electrons are removed from $CuO_2$ planes) some of these "unhappy" electrons loose their prejudice of being localized on copper sites\footnote{Anderson pointed out that the amount of distortion of the octahedron for $Cu^{++}$ (one hole and one unpaired electron in $3d_{x^2-y^2}$ orbitals) and for $Cu^{+++}$ (two holes in $3d_{x^2-y^2}$ orbitals) states is roughly the same and the doped hole can become mobile without taking the "burden" of a lattice distortion along with it, i.e., when a doped hole migrates from one site to another, it does not carry along with it a distortion (lattice distortion) of the octahedron. Contrary to it $Cu^{+}$ (two electrons in $3d_{x^2-y^2}$ orbitals) state is of spherical symmetry and involve a distortion of the octahedron and not very supportive for easy conduction as the mobile hole has to carry a lattice distortion along with it\cite{ander1}. This might be the reason why, in the electron doped side, AFM order is so much extended in the phase diagram and SC is so much weaker as compared to that in the hole doped side.}. How does this happen?  {\it It is not only "unhappy" electrons on copper sites which are affected by hole doping, populations on oxygen sites do also change}\cite{gau}. Conduction becomes possible (doped holes are mobile objects). In this conducting "metal" short range AFM correlations are dominant in the underdoped regime and survive even beyond the optimal doping regime so long as superconductivity is present (Figure 2). 
\begin{figure}[!h]
\begin{center}
\includegraphics[height=4cm]{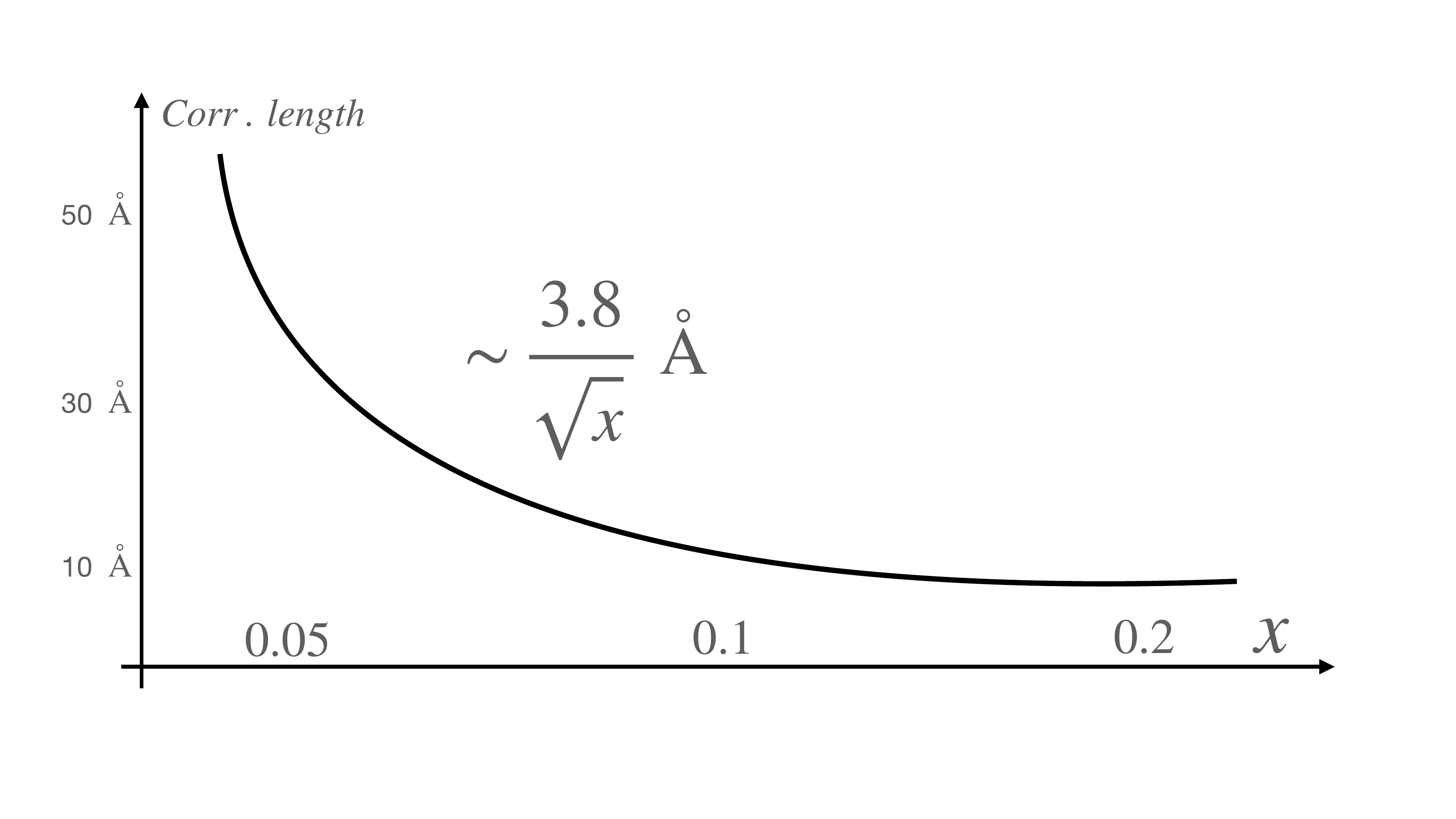}
\caption{AFM correlation length from neutron scattering in $La_{2-x}Sr_xCuO_4$\cite{dago}.}
\end{center}
\end{figure}
These short range AFM correlations (mediated through superexchange interaction) are some sort of a relics of the full fledged AFM order that existed at zero doping.  Cuprates are "naturally paired", that is,  there exists short-ranged AFM correlations which evolve in time (a patch of local AFM order extending over several lattice sites appears at a given instant of time, and then it dies aways, and it appears again at a different location, and so on. {\it This process is incessant}). This state is a fundamental state out of which superconductivity emerges. Superconductivity in cuprates require two essential ingredients: presence of mobile states and presence of short ranged AFM correlations. Do the doped holes change the average local moments on copper sites? The answer is yes. They diminish them and can provide answers to neutron scattering experiments.  Barzykin and Pines argued based on extensive experimental analysis that both local degrees-of-freedom that is the cause of magnetic moments (spin liquid component) and mobile degrees-of-freedom that is the cause of conduction (Fermi liquid component) exist in underdopped and optimally doped cuprates\cite{bp}. They argued that the strength of effect exchange interaction in the spin liquid component weakens by increasing doping:
\beq
J_{eff}(x) = J f(x),~~~f(x)= 1-x/0.2.
\eeq
That is, spin liquid component weakens with increasing doping whereas Fermi liquid component becomes stronger with increasing doping. But beyond optimal doping or more precisely beyond a critical doping of $p>p^*\simeq 0.19$ something very drastic happens.

In 2016, an important discovery\cite{taillefer1} highlighted this drastic change of behaviour across the pseudogap quantum critcal point at $p^* \simeq 0.19$. Hall effect measurements were done on YBCO. These measurements showed that in the underdoped regime below $p<0.1$ the number of carriers per Cu atom is simply equal to $p$ and in the overdoped side ($p>0.19$) this number is $1+p$. There is a "sudden" transition somewhere between these two dopings.  As the doping is reduced from overdoping to underdoping, this change in the number of carriers begins at $p=p^*$ and it gradually reduces towards $p$ holes per Cu atom at and below $p=0.1$ as doping is reduced from overdoping to underdoping (refer Figure 3). 

\begin{figure}[!h]
\begin{center}
\includegraphics[height=8cm]{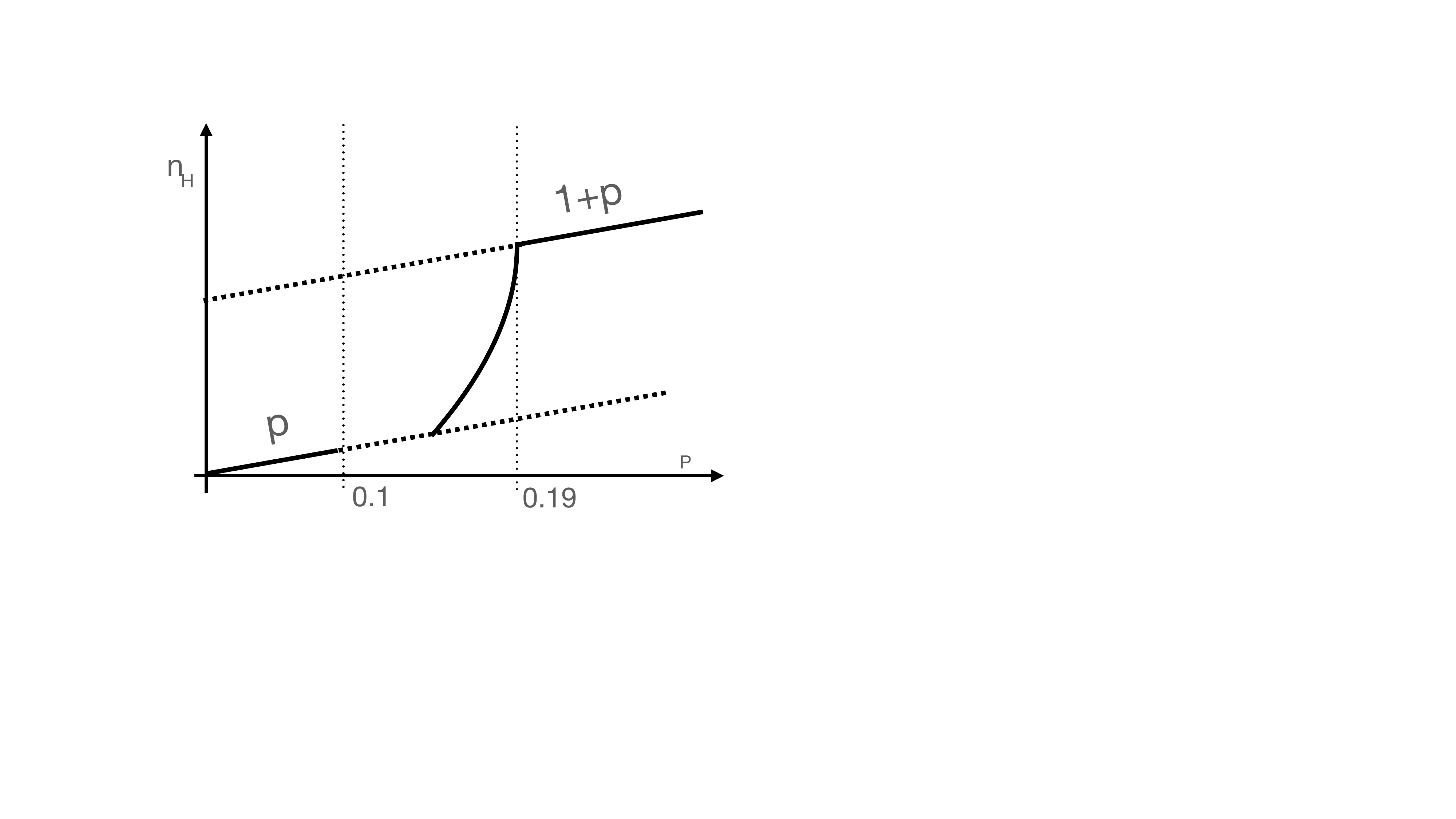}
\caption{$1+p$ to $p$ transition as doping is reduced through $p^*$.}
\end{center}
\end{figure}

Something drastic happens to Anderson's unhappy electrons at around $p^*$ where these are at the verge of itineracy.  Below $p^* ~(p<p^*)$ they are localized and above $p^*$ they are mobile. At $p=p^*$ they are not able to "decide" whether to localize or to be itinerant.  We call this a "split personality state" of these "unhappy" electrons. They become mobile for $p>p^*$ giving total carrier number as $1+p$ per copper atom as has been discussed by Louis Taillefer and collaborators\cite{taillefer1} and below $p^*$ these unhappy electrons gradually begin to localize on copper sites, and the carrier number is equal to the number of doped holes. At  this critical doping Fermi surface volume changes from a bigger volume of $1+p$ carriers per copper atom to a smaller volume of $p$ carriers per copper atom as the doping is reduced through $p^*$. Experiments cannot be done at zero temperature and extrapolation shows that this transformation occurs at $p^*$ (Figure 3). But at finite temperature this transformation through $p^*$ is not sharp but is a cross-over\cite{taillefer1}.  The question is:  What is the role played by this transition in the mechanism of superconductivity? And a connected question is: Is  $1+p$ to $p$ transition around $p^*$  the root cause of the higher temperature pseudogap (at which the static magnetic susceptibility shows a maximum)? Answer to the latter  question, with the available evidence, seems to be affirmative. Static magnetic susceptibility shows a maximum when electrons are at the verge of itineracy or when they are in a state not able to "decide" whether to localized or be itinerant. In fact, generalized susceptibilities show divergences at critical points.   Such a maximum has been observed in the static magnetic susceptibility of underdoped cuprates. At higher temperatures, thermal effects broaden the peak.

Let us ask the first question in a slightly different way: Are local moments on copper sites must for superconductivity? How does the change of character beyond the critical doping of $p^*$ affects superconductivity? Is it behind the destruction of superconducting dome at overdopping? And why does $T_c$ maximizes around $p^*$ where the electrons are in "split personality state". In fact $T_c$ is maximum around $p\simeq 0.16$ not around $p = 0.19$. Is this observation very crucial and fundamental? If this is a crucial observation then SC is not a phenomena ONLY governed by what happens at $p^*$. It is maximum when "Goldilocks" situations arises. {\it "Goldilocks" situation occurs when the system achieves the best compromise between the growing number of carriers (good for superfluid density) as doping is increased and simultaneously decreasing short ranged AFM correlations (bad for superconductivity).}\footnote{For a discussion and problems of theories based on these lines, refer to\cite{nav1}.}  At overdoping when AFM correlations become non-existent  SC disappears. But some of the heavy Fermion superconductors (HFSC) portray a different picture. In these systems SC order emerges when long range magnetic order is suppressed (by pressure or chemical doping). SC dome appears around the magnetic quantum critical point. At a sufficiently high pressure (or when system is tuned much away from QCP) SC disappears as magnetic correlations weaken or disappear. How do we understand the difference between what happens at $p\simeq 0.16$ and at $p\simeq 0.19$?

Let us ask a connected question: why does superconductivity occur at overdopping (just beyond $p>0.19$ and below the end point of the dome) when all the carriers become mobile?  The very fact that superconductivity is present at overdopping means that antiferromagnetic correlations still present at over doping (but below the doping level at which SC dome ends. Refer to Figure 4). They disappear only at overdopping when there is no SC and when there is $T^2$ behaviour of DC resistivity. $T^2$ behaviour of DC resistivity is a clear signature that short ranged AFM correlations become sub-dominant in this overdoped regime, and the situation is well described by Fermi liquid theory\cite{nav1}.

Let us change gears and discuss about the most vexed issue of the pseudogapped (PG) state of cuprates.  Two lines are generally drawn for the pseudogapped state in underdopped Cuprates: one starting at high temperatures (approximately around 1200 K at zero doping) and then approximately linearly going down with increasing doping and meeting the doping axis at a critical doing of $p=0.19$ in the phase diagram (refer to Figure 4), and the second line is generally drawn starting from a lower temperature of 300 K or 400 K (depending upon the cuprate family) at zero doping, just above the Neel ordering temperature. This line also decreases linearly with increasing doping and ends at the quantum critical point ($p=0.19$) or in some scenarios becoming tangent to the dome at overdoping\cite{yves1}. These lines are cross overs, and we denote the upper line (high temperature PG) with $T^*$\footnote{The upper crossover has been called an energy scale $E_{pg}$ (not a temperature scale) and $T^* = \frac{E_{pg}}{k_B}$. This point is stressed by Tallon and collaborators\cite{tallon}.} and the lower line (low temperature PG) with $T^{**}$.

The mechanism behind the high temperature PG is connected with Taillefer's transition of $p$ to $1+p$ holes per copper atom when doping is increased through the quantum critical point. That is, it is connected with de-localization to localization transition of Anderson's "unhappy" electrons.  When temperature $T$ is reduced through $T^*$ boundary,  quasi-local moments appear on copper sites.  This high temperature PG ($T^*$ line) is a single plane phenomena. It is a direct product of Mott type correlations which begins when local moments appear on copper sites below $T^*$, and Mott correlations remain active even at finite dopings (for $p<p^*$).  In the heat capacity experiments gap is observed to open at and below the critical doping level $p^*$\cite{tallon}. The low temperature part of it is connected with the Fermi surface volume reduction when doping is reduced through $p^*$ (refer to Figure 3). Also, it is the same temperature scale (generally marked as $T_{max}$) at which static magnetic susceptibility shows a maximum.

\begin{figure}[!h]
\begin{center}
\includegraphics[height=7cm]{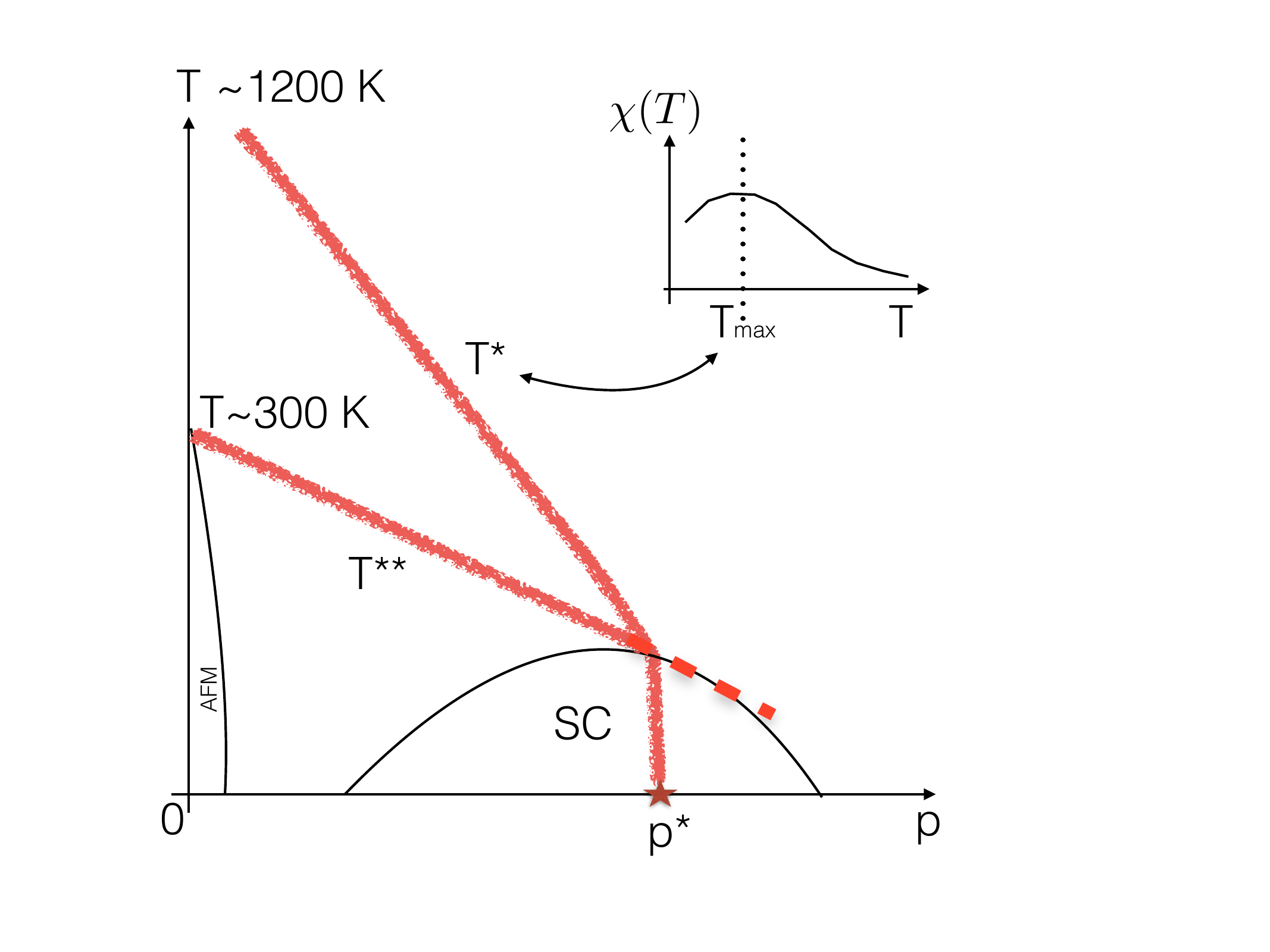}
\caption{The high temperature ($T^*$) and the low temperature ($T^{**}$) pseudogaps in the the phase diagram.}
\end{center}
\end{figure}

The second line (low temperature PG) is very enigmatic. Experimental signatures of it are: (1) Reduction in $\frac{1}{T_1^{63}T}$ relaxation rate below $T^{**}$; (2) reduction (down-turn) in in-plane DC resistivity; (3) upturn in c-axis resistivity etc. For a review refer to\cite{timuskstatt, norman}.

Various mechanisms has been proposed to address this low temperature PG: (1) pre-formed  pairs and related ideas\cite{yves1, yves2, huf, mohit, curty};  (2) competing orders\cite{sudip}; (3) It could be an out-come of inter-plane "coherence" or 3D effects beyond the single CuO2 planes.  At zero doping in-plane local moments on Cu sites appear around $\sim 1200~K$ and they exhibit AFM only around $\sim 300 ~K$ when inter-plane correlations set-in (3D effects). The question that we ask is this: Whether these zero doping inter-plane effects vanish when some small doping is made in the CuO2 planes? \footnote{True nature of this interlayer coupling between fluctuating short range AFM order in one $CuO_2$ layer with another $CuO_2$ layer is elusive, but it could be that a newly born patch of local short range AFM order in one $CuO_2$ layer aligns a similar patch of local AFM order in adjacent $CuO_2$ layers just above or below it. And this interlayer correlation is transient in time (coming and going with time). What is the effect of pressure on $T^{**}$?};  (4) $T^{**}$ and associated spin damping phenomena\cite{jorg}; (5)  Nematicity: $T^{**}$ it might be that degeneracy between oxygen $p$ orbitals is lifted that leads to nematicity. Recently, the connection between nematicity and lifting of degeneracy between oxygen $p$ orbitals has been pointed out (energy difference between $p_x$ and $p_y$ orbitals is found to be roughly $50~meV$)\cite{sou,shu}.

Regarding nematicity, in 2019, an investigation\cite{sou} hinted towards a link between vetigial density waves (DW) and vestigial nematicity associated with the PG state. A conundrum was presented: Density wave state (DW with $Q\ne0$) that could produce an energy gap in EDOS is only detectable at temperatures much lower than that of the PG temperature. No new additional energy gap is opened at the on-set of DWs. On the other hand, nematic state ($Q=0$) that appears at the PG temperature is not capable opening the gap! So what is the mechanism behind the gap opening? It is argued that some sort of the relics of the DWs (vestigial DWs) are responsible\footnote{Here vistigial nematicity means both $C_4$ symmetry breaking (remnants of DW order) and rotational symmetry breaking (remnants of AFM order).}. However, the resolution remains wide open. Trouble starts when we try to reconcile the above conundrum with the wisdom gained (in early years of 2000) by David Pines and collaborators regarding the lower PG where it is associated with some sort of spin damping phenomena below $T^{**}$\cite{norman, jorg}. Spins become localized below $T^*$ and exhibit Mott type correlations. At a still lower temperatures, below $T^{**}$, something again happens to spin system (sharp drop in NMR relaxation rate). How do we achieve a comprehensive understanding? Clearly,  these are very hard questions and provoke compound by compound analysis. Resolution among these scenarios is much needed. It puts a heavy demand of a very careful analysis of literature which has grown tremendously large over the years. A detailed and critical discussion will be presented in a forthcoming article:\footnote{"The pseudogap debate: current status", manuscript under preparation. What is universal and what is compound specific will be distinguished and analysed.}

With this, we end our brief review/perspective. Coming back to the title of this article, Anderson's "unhappy" electrons start their journey as being localized at zero doping and becoming quasi-local as doping is increased. Doped holes stay mobile but they diminish the magnitude of local moments.  Short ranged AFM correlations remain active. From the overdoped side as doping is reduced, PG opens when some of them become quasi-localized at $Cu$ sites below $p^*$. This quasi-localization triggers Mott type correlations and that open up a gap (the high temperature pseudogap at $T^*$). Mechanism behind the lower PG at $T^{**}$ remains elusive.

\section*{Acknowledgments}
Author would like to thank Yves Noat, Alain Mauger, and William Sacks for illuminating discussions regarding the pseudogapped state.
\section*{References}



\end{document}